\documentclass[twocolumn,showpacs,preprintnumbers,amsmath,amssymb]{revtex4}

\usepackage{graphicx}

\begin{document}

\title{The single-particle density matrix and 
the momentum distribution of dark "solitons" 
in a Tonks-Girardeau gas}

\author{H. Buljan,$^{1}$ K. Lelas,$^{1}$ R. Pezer,$^{2}$ and M. Jablan$^{1}$}
\affiliation{$^{1}$Department of Physics, 
University of Zagreb, PP 332, Zagreb, Croatia}
\affiliation{$^{2}$Faculty of Metallurgy, University of Zagreb, 
Aleja narodnih heroja 3, Sisak, Croatia}
\date{\today}


\begin{abstract}
We study the reduced single-particle density matrix (RSPDM), 
the momentum distribution, natural orbitals and their occupancies, 
of dark "soliton" (DS) states in a Tonks-Girardeau gas. 
DS states are specially tailored excited many-body eigenstates, 
which have a dark solitonic notch in their single-particle density. 
The momentum distribution of DS states has a characteristic shape 
with two sharp spikes. 
We find that the two spikes arise due to the high degree
of correlation observed within the RSPDM between the mirror points 
($x$ and $-x$) with respect to the dark notch at $x=0$;
the correlations oscillate rather than decay 
as the points $x$ and $-x$ are being separated. 
\end{abstract}

\maketitle

\section{Introduction}

Exactly solvable models have the possibility of providing 
important insight into the quantum many-body physics beyond 
various approximation schemes. 
Two of such models, the Tonks-Girardeau \cite{Girardeau1960} and the Lieb-Liniger model 
\cite{Lieb1963}, which describe interacting Bose gases in one dimension (1D), 
have drawn considerable attention in recent years with the developments 
of the experimental techniques for tightly confining atoms in 
effectively 1D atomic waveguides \cite{OneD,Kinoshita2004,Paredes2004,Kinoshita2006}.
The Lieb-Liniger model (LL) describes a system of bosons
interacting via two-body $\delta$-function interactions \cite{Lieb1963}. 
The Tonks-Girardeau (TG) model corresponds to  
infinitely repulsive ("impenetrable core") bosons in 1D \cite{Girardeau1960,LL-TG};
this model is exactly solvable via Fermi-Bose mapping, which 
relates the TG gas to a system of noninteracting spinless fermions 
in 1D \cite{Girardeau1960}. A study of atomic scattering 
for atoms confined transversally in an atomic waveguide 
has lead to a suggestion for the experimental observation 
of a TG gas \cite{Olshanii}; such atomic systems enter the TG regime 
at low temperatures, low linear densities, and strong effective 
interactions \cite{Olshanii,Petrov,Dunjko}. 
The experimental realization of the TG model was 
reported in two experiments from 2004 \cite{Kinoshita2004,Paredes2004}. 
Moreover, nonequilibrium dynamics of the 1D interacting Bose gases
including the TG regime has been recently experimentally studied within 
the context of relaxation to an equilibrium \cite{Kinoshita2004}. 
Within this paper we analyze the reduced single-particle density 
matrix (RSPDM) and related observables of certain specially tailored excited 
eigenstates of the TG gas, which are also referred to as dark 
"soliton" (DS) states \cite{Girardeau2000,Busch2003,Buljan2006}.

Dark solitons are fundamental nonlinear excitations. Within the 
context of interacting Bose gases, they were mainly studied in the 
regime of weak repulsive interactions \cite{ExpDark,Dum1998,
Busch2000,Muryshev2002} were mean-field theories
[e.g., the Gross-Pitaevskii theory, which employs the nonlinear 
Schr\" odinger equation (NLSE)] are applicable. 
In the regime of strong repulsive interactions in quasi-1D geometry, 
dark solitons were also studied by using NLSE with a quintic 
nonlinear term \cite{Kolomeisky2000,Frantzeskakis2004,Ogren2005}. 
In Ref. \cite{Girardeau2000}, Girardeau and Wright have studied 
the concept of dark solitons within the exactly solvable 
TG model; they found specially tailored excited many-body eigenstates of the 
TG gas on the ring (DS states), with a dark notch in their single-particle 
density, which is similar to the dark notch of nonlinear dark-solitons. 
The dynamics of such excitations in a TG gas was studied by Busch and Huyet 
\cite{Busch2003} in a harmonic trap. 
Recently, a scheme based on parity selective filtering ("evaporation") 
of a many-body wave function was suggested \cite{Buljan2006} as a 
candidate for the experimental observation of DS states. 
However, to the best of our knowledge, the momentum distribution, the 
RSPDM, natural orbitals (NOs) and their occupancies, have not been studied 
yet for DS states. These quantities are 
important for the better understanding of DS states, 
but may also be necessary ingredients for their experimental detection, 
which provides motivation for this study.

The calculation of correlation functions (such as the RSPDM) 
for 1D Bose gases \cite{Lenard1964,Creamer1981,Girardeau2001,Minguzzi2002,
Cazallilla2002,Olshanii2003,Papenbrock2003,Forrester2003,Gangardt2003,
Astrakharchik2003,Rigol2004,Gangardt2004,Berman2004,Rigol2005,
Minguzzi2005,Brand2005,Forrester2006,Gangardt2006,Rigol2006,Pezer2007,
Deuretzbacher2007,Caux2007,Lin2007} 
from the many-body wave functions 
\cite{Girardeau1960,Lieb1963,Girardeau2000,Muga1998,Sakmann2005,Batchelor2005} yields 
important physical information (such as the momentum distribution) 
on the state of the system. 
Within the TG model, the RSPDM and the 
momentum distribution have been studied in the continuous 
\cite{Lenard1964,Girardeau2001,Minguzzi2002,Papenbrock2003,
Forrester2003,Gangardt2004,Minguzzi2005,Pezer2007,Lin2007}, and discrete (lattice) case 
\cite{Rigol2004,Rigol2005,Gangardt2006,Rigol2006,Cazalilla2004}, 
both for the static 
\cite{Lenard1964,Girardeau2001,Minguzzi2002,Papenbrock2003,
Forrester2003,Rigol2004,Gangardt2004,Gangardt2006,Lin2007} 
and time-dependent problems 
\cite{Rigol2005,Minguzzi2005,Rigol2006,Pezer2007}. 
In the stationary case, most studies consider the ground state 
properties of the TG gas. 
The momentum distribution for the ground state of the 
TG gas on the ring has a spike at $k=0$,  
$n_B(k)\propto |k|^{-1/2}$ \cite{Lenard1964}. 
In both the harmonic confinement \cite{Minguzzi2002,Olshanii2003} 
and on the ring \cite{Olshanii2003}, the TG ground state 
momentum distribution decays as a power law $n_B(k)\propto k^{-4}$; 
in Ref. \cite{Olshanii2003} it has been pointed out that 
$k^{-4}$-decay is also valid for the LL gas (for any 
strength of the interaction). 
These ground states of the TG gas are not Bose condensed 
\cite{Lenard1964,Forrester2003}, which is evident from 
the fact that the occupancy of the leading 
natural orbital scales as $\sqrt{N}$ for large $N$
\cite{Forrester2003,Papenbrock2003}. 
In the box-confinement, the momentum distribution of a TG gas 
has been studied by generalizing the Haldane's harmonic-fluid approach
\cite{Cazallilla2002}. 
Besides for the ground states, the momentum distribution 
has been analyzed in time-dependent problems including 
irregular dynamics on the ring \cite{Berman2004}, dynamics 
in the harmonic potential with time dependent frequency \cite{Minguzzi2005}, 
and in a periodic potential in the context of many-body 
Bragg reflections \cite{Pezer2007}. 
A number of interesting results for time-dependent problems have been 
recently obtained within the discrete lattice model including 
fermionization of the momentum distribution during 1D free expansion 
\cite{Rigol2005}, and relaxation to a steady state carrying memory 
of initial conditions \cite{Rigol2006}.

The correlation functions for TG and LL models were studied by 
using various analytical and numerical methods
\cite{Lenard1964,Creamer1981,Girardeau2001,Minguzzi2002,
Cazallilla2002,Olshanii2003,Papenbrock2003,Forrester2003,Gangardt2003,
Astrakharchik2003,Rigol2004,Gangardt2004,Berman2004,Rigol2005,
Minguzzi2005,Brand2005,Forrester2006,Gangardt2006,Rigol2006,Pezer2007,
Deuretzbacher2007,Caux2007,Lin2007}.
The formula that was derived and employed in Ref. \cite{Pezer2007} 
allows efficient and exact numerical calculation of the RSPDM 
for the TG gas in versatile states (ground state, excited eigenstates, 
time-evolving states \cite{Pezer2007}), and for a fairly large number of particles. 
We find it suitable for this study of DS states.

Here we numerically calculate the RSPDM correlations, natural orbitals and 
their occupancies, and the momentum distribution of DS states. 
We find that these excited eigenstates 
of a TG gas have characteristic shape of the momentum 
distribution with two sharp spikes. 
The two sharp spikes arise due to the high degree
of correlation observed within the RSPDM between the mirror points, 
$x$ and $-x$, with respect to the dark notch at $x=0$;
interestingly, the correlations oscillate rather than decay 
as the points $x$ and $-x$ are being separated.

\section{The model}

We study a system of $N$ identical Bose particles in 1D space,
which experience an external potential $V(x)$. 
The bosons interact with impenetrable pointlike interactions 
\cite{Girardeau1960}, which is most conveniently represented 
as a subsidiary condition on the many-body wave function 
\cite{Girardeau1960}:

\begin{equation}
\psi_B(x_1,x_2,\ldots,x_N,t)=0\ \mbox{if}\ x_i=x_j
\end{equation}
for any $i\neq j$. 
Besides this condition, $\psi_B$ obeys the Schr\" odinger equation 

\begin{equation}
i \frac{\partial \psi_B}{\partial t}=
\sum_{j=1}^{N} \left[ -\frac{\partial^2 }{\partial x_j^2}
+ V(x_j) \right] \psi_B;
\end{equation}
here we use dimensionless units as in Ref. \cite{Buljan2006}, i.e., 
$x=X/X_0$, $t=T/T_0$, and $V(x)=U(X)/E_0$, where 
$X$ and $T$ are space and time variables in physical units, 
$X_0$ is an arbitrary spatial length-scale (e.g., $X_0=1\ \mu$m), 
which sets the time-scale $T_0=2mX_0^2/\hbar$, and energy-scale 
$E_0=\hbar^2/(2mX_0^2)$; $m$ denotes particle mass, and 
$U(X)$ is the potential in physical units. 
The wave functions are normalized as 
$\int dx_1\ldots dx_N |\psi_B(x_1,x_2,\ldots,x_N,t)|^2=1$.

The solution of this system may be written in compact form via the 
famous Fermi-Bose mapping, which relates the TG bosonic wave function 
$\psi_B$ to an antisymmetric many-body wave function $\psi_F$ 
describing a system of noninteracting spinless fermions 
in 1D \cite{Girardeau1960}:

\begin{equation}
\psi_B = 
A(x_1,\ldots,x_N) \psi_F(x_1,x_2,\ldots,x_N,t).
\label{mapFB}
\end{equation}
Here

\begin{equation}
A=\Pi_{1\leq i < j\leq N} \mbox{sgn}(x_i-x_j)
\label{unitA}
\end{equation}
is a "unit antisymmetric function" \cite{Girardeau1960},
which ensures that $\psi_B$ has proper bosonic symmetry 
under the exchange of two bosons. The fermionic wave function
$\psi_F$ is compactly written in a form of the Slater determinant,

\begin{equation}
\psi_F(x_1,\ldots,x_N,t)=
\frac{1}{\sqrt{N!}} \det_{m,j=1}^{N} [\psi_m(x_j,t)], 
\label{psiF}
\end{equation} 
where $\psi_m(x,t)$ denote $N$ orthonormal SP 
wave functions obeying a set of uncoupled single-particle (SP)
Schr\" odinger equations 

\begin{equation}
i\frac{\partial \psi_m}{\partial t}=
\left [ - \frac{\partial^2 }{\partial x^2}+
V(x) \right ] \psi_m(x,t), \ m=1,\ldots,N.
\label{master}
\end{equation}
Equations (\ref{mapFB})-(\ref{master}) prescribe construction of
the many-body wave function describing the TG gas in an external 
potential $V(x)$, both in the static \cite{Girardeau1960} and 
time-dependent case \cite{Girardeau2000}. 
The eigenstates of the TG system are

\begin{equation}
\psi_B(x_1,\ldots,x_N)=A(x_1,\ldots,x_N)
\frac{1}{\sqrt{N!}} \det_{m,j=1}^{N} [\phi_m(x_j)], 
\label{psiBeig}
\end{equation} 
where $\phi_m(x)$ are single-particle eigenstates 
for the potential $V(x)$. In the rest of the paper we will 
discuss the eigenstates of the TG system and their observables; 
hence, we drop the time-variable from subsequent notation.

The many-body wave function $\psi_B$ fully describes 
the state of the system. However, its form does 
not transparently yield physical information related to 
many important observables (e.g., the momentum distribution). 
The expectation values of one-body observables 
are readily obtained from the RSPDM, defined as 

\begin{eqnarray}
\rho_{B}(x,y) & = &  N \int \!\! dx_2\ldots dx_N \, \psi_B(x,x_2,\ldots,x_N)^*
\nonumber \\
&& \times \psi_B(y,x_2,\ldots,x_N).
\end{eqnarray} 
The observables of great interest are the 
SP $x$-density $\rho_{B}(x,x)=\sum_{m=1}^{N}|\phi_m(x)|^2$, and the momentum 
distribution \cite{Lenard1964}:

\begin{equation}
n_B(k) = \frac{1}{2\pi}\int \!\! dx dy \, e^{i k(x-y)}\rho_{B}(x,y). 
\label{MDformula}
\end{equation} 
The SP density $\rho_{B}(x,x)$ is identical for 
the TG gas and the noninteracting Fermi gas \cite{Girardeau1960},
however, the momentum distributions of the two systems considerably 
differ \cite{Lenard1964}.

A concept that is very useful for the understanding 
of the many-body systems is that of natural orbitals (NOs). 
The NOs $\Phi_i(x)$ are eigenfunctions of 
the RSPDM, 

\begin{equation}
\int \!\! dx\, \rho_{B}(x,y) \, \Phi_i (x) =
\lambda_i \, \Phi_i (y), \quad i=1,2,\ldots,
\end{equation}
where $\lambda_i$ are the corresponding eigenvalues;
the RSPDM is diagonal in the basis of NOs,

\begin{equation}
\rho_{B}(x,y) = \sum_{i=1}^{\infty} 
\lambda_i \Phi_i^* (x) \Phi_i (y).
\end{equation}
The NOs can be interpreted as effective SP states
occupied by the bosons, where $\lambda_i$ 
represents the occupancy of the corresponding NO \cite{Girardeau2001}. 
The sum of the Fourier power spectra of the NOs 
is the momentum distribution:

\begin{eqnarray}
n_B(k) = \sum_{i=1}^{\infty} 
\lambda_i \tilde\Phi_i^* (k) \tilde\Phi_i (k),
\label{BMDNOs}
\end{eqnarray} 
where $\tilde\Phi_i (k)$ is the Fourier transform of $\Phi_i (x)$.

The RSPDM of the noninteracting fermionic system on the 
Fermi side of the mapping is 

\begin{equation}
\rho_{F}(x,y)=\sum_{m=1}^{N}\phi_m^*(x)\phi_m(y);
\end{equation}
evidently, the SP eigenstates $\phi_m(x_j)$ 
are NOs of the fermionic system, with occupancy unity \cite{Girardeau2001}. 
The fermionic momentum distribution is 

\begin{eqnarray}
n_F(k) = \sum_{m=1}^{N} \tilde\phi_m^* (k) \tilde\phi_m (k),
\label{FMDNOs}
\end{eqnarray} 
where $\tilde\phi_m (k)$ is the Fourier transform of $\phi_m (x)$.

The calculation of the TG momentum distribution is preceded by a 
calculation of $\rho_B(x,y)$, which we conduct according 
to the method described in Ref. \cite{Pezer2007}. 
If the RSPDM is expressed in terms of the SP 
eigenstates $\phi_m$ as 

\begin{equation}
\rho_{B}(x,y)=\sum_{i,j=1}^{N}
\phi^{*}_{i}(x)A_{ij}(x,y)\phi_{j}(y),
\label{expansion}
\end{equation}
it can be shown that the $N\times N$ matrix ${\mathbf A}(x,y)=\{ A_{ij}(x,y) \}$ 
has the form

\begin{equation}
{\mathbf A}(x,y)=  ({\mathbf P}^{-1})^{T} \det {\mathbf P},
\label{formulA}
\end{equation} 
where the entries of the matrix ${\mathbf P}$ are 
$P_{ij}(x,y)=\delta_{ij}-2\int_{x}^{y}dx' \phi_{i}^{*}(x')\phi_{j}(x')$
($x<y$ without loss of generality) \cite{Pezer2007}. 
Formulas (\ref{expansion}) and (\ref{formulA}) enable fast numerical 
calculation of the RSPDM (and related quantities) for dark "soliton" states.

\section{DS states on the ring}

Within this section we analyze the RSPDM, the momentum distribution,
NOs and their occupancies for excited eigenstates of a TG gas on the 
ring of length $L$; in other words, external potential is zero, 
$x$-space is $x \in [-L/2,L/2]$, and periodic boundary conditions are 
imposed. The many-body eigenstates of the TG gas are 
constructed from the SP eigenstates of the system 
via Eq. (\ref{psiBeig}). Th SP eigenstates for the 
ring geometry are plane waves $\sqrt{1/L}e^{ik_m x}$, with 
SP energy $k_m^2$; here $k_m=2\pi m/L$, and $m$ is integer
\cite{OddN}. 
Apparently, the eigenstates $\sqrt{1/L}e^{ik_m x}$ and 
$\sqrt{1/L}e^{-ik_m x}$ are degenerate. 
This degeneracy in the SP eigenstates induces [via Eq. (\ref{psiBeig})] 
degeneracy of the TG many-body excited eigenstates. 
One particular subspace of degenerate eigenstates (DEs) 
is spanned with 

\begin{equation}
\phi_m(x) =\frac{1}{\sqrt{L}}[a_{m}^{-}e^{-ik_m x}+a_{m}^{+}e^{ik_m x}] 
\end{equation}
where $|a_{m}^{-}|^2+|a_{m}^{+}|^2=1$, and $m=1,\ldots,N$; 
the corresponding many-body eigenstates are 

\begin{eqnarray}
\psi_{DE} & = & 
A(x_1,\ldots,x_N)
L^{-\frac{N}{2}} 
\times
\nonumber \\
&&\det_{j,m=1}^{N}[a_{m}^{-}e^{-ik_m x_j}+a_{m}^{+}e^{ik_m x_j}].
\label{DE}
\end{eqnarray}
Intuition suggests that, although these states are 
degenerate, some of the corresponding observables,
such as the SP density in $x$-space, the momentum distribution, 
spatial coherence or entropy, could be quite different from 
one eigenstate to another depending on their internal symmetry, 
which is designated by the choice of the 
coefficients $a_{m}^{-}$ and $a_{m}^{+}$.

In Ref. \cite{Girardeau2000}, Girardeau and Wright have 
pointed out that if one constructs excited many-body eigenstates 
of the TG gas on the ring as 

\begin{eqnarray}
\psi_{DS} & = & 
A(x_1,\ldots,x_N)
\left ( \frac{2}{L} \right)^{\frac{N}{2}} \times
\nonumber \\
&&\det_{j,m=1}^{N}[\sin k_m x_j],
\label{psiDS}
\end{eqnarray}
that is, if one chooses the coefficients as 
$a_{m}^{-}=i/\sqrt{2}$ and $a_{m}^{+}=-i/\sqrt{2}$, the 
SP density of these many-body eigenstates \cite{Girardeau2000},

\begin{equation}
\rho_{DS}(x,x)=\frac{N+1}{L}-
\frac{\sin(\frac{(N+1)2\pi x}{L})\cos(\frac{N2\pi x}{L})}
{L\sin(\frac{2\pi x}{L})},
\label{rhoDSxx}
\end{equation}
has the structure closely resembling dark solitons \cite{Girardeau2000} 
(hence the notation $\psi_{DS}$ for the many-body wave function, and 
analogously for related observables below).  
The structure of these excited eigenstates is somewhat artificial 
because on the fermionic side of the mapping, these states 
correspond to noninteracting fermions being placed solely within 
the {\em odd} SP eigenstates $\sin k_m x$. Nevertheless, such states can be 
excited by filtering of the many-body wave function \cite{Buljan2006}. 

\begin{figure}
\begin{center}
\includegraphics[width=0.4 \textwidth ]{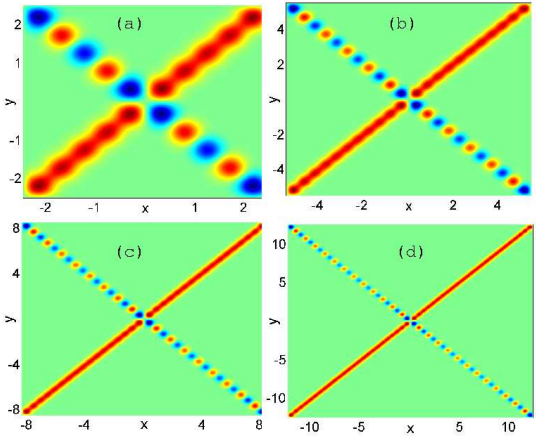}
\caption{ \label{rhodark}
(color online) The RSPDM of dark "soliton" states, $\rho_{DS}(x,y)$, 
for $N=5$ (a), 11 (b), 17 (c), and 25 (d). 
}
\end{center}
\end{figure}

Let us utilize the procedure outlined in Sec. II 
to calculate the RSPDM, and related one-body observables
for DS states [Eq. (\ref{psiDS})]. 
It is straightforward to calculate the entries of the matrix 
${\mathbf P}={\mathbf 1}-{\mathbf Q}$ [ see Eq. (\ref{formulA})], where 

\begin{eqnarray}
Q_{ij} & = & \frac{\sin (2(i+j)\pi x/L)}{(i+j)\pi}-\frac{\sin(2(i-j)\pi x/L)}{(i-j)\pi}
\nonumber \\
&& -\frac{\sin(2(i+j)\pi y/L)}{(i+j)\pi}+\frac{\sin(2(i-j)\pi y/L)}{(i-j)\pi},\ i\neq j;
\nonumber \\
Q_{ii} & = &  -2\frac{x-y}{L}
+\frac{ \sin (\frac{4 i \pi  x}{L})}{2 i \pi }
-\frac{ \sin (\frac{4 i \pi  y}{L})}{2 i \pi };
\end{eqnarray}
for $i,j=1,\ldots,N$. 
As for the inverse of the matrix ${\mathbf P}$, and consequently 
the RSPDM, we were able to find its analytical form up to $N=7$ 
by using {\em Mathematica}. 
However, for larger $N$ we resorted to numerical calculations. 
It is straightforward to see that the RSPDMs of two DS states, 
for two different values of $L$, say $L_1$ and $L_2$, are connected
by a simple scaling,

\begin{equation}
L_1 \rho_{DS,L_1}(x L_1,y L_1)
=
L_2 \rho_{DS,L_2}(x L_2,y L_2),
\end{equation}
where $x,y\in [-\frac{1}{2},\frac{1}{2}]$;
thus, it is sufficient to calculate it for just one value of $L$.
In what follows, without loosing any generality, we choose $N=L$. 

\begin{figure}
\begin{center}
\includegraphics[width=0.4 \textwidth ]{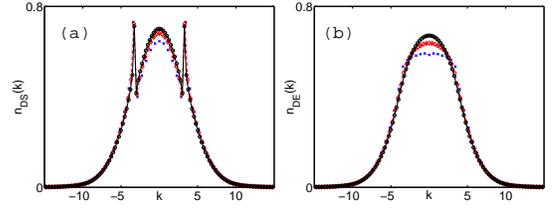}
\caption{ \label{MDdark}
(color online) The momentum distributions corresponding to 
DS states (a), and to degenerate eigenstates with 
randomly chosen phases (b); figures are shown for $N=11$ (x-symbols, 
blue dotted line), $17$ (diamonds, red dashed line), and $25$ 
(circles, solid black line). 
}
\end{center}
\end{figure}

Figure \ref{rhodark} displays contour plots of $\rho_{DS}(x,y)$ 
for $N=5,11,17$ and $25$. We clearly see a characteristic pattern for 
each value of $N$: The RSPDMs are largest close to the diagonal, with 
oscillations following the $x$-space density from 
Eq. (\ref{rhoDSxx}). However, there are strong correlations 
along the line $x=-y$ indicating coherence between mirror 
points $x$ and $-x$ around the DS center (at $x=0$).

Figure \ref{MDdark}(a) displays the momentum distribution 
$n_{DS}(k)$ of DS states for $N=11,17$, and $25$. 
All momentum distributions for the ring geometry are 
normalized as $\sum_{k_m} n_B(k_m)=N$ 
(the SP momentum values $k_m$ are discrete in the ring geometry). 
The momentum distributions have a characteristic shape 
with a smooth hump close to the origin ($k=0$), and with two 
sharp spikes which are located at $\pm k_{peak}=\pm\sum_{m=1}^{N}k_m/N =\pm \pi (N+1) /L$; 
the spikes indicate that there is high probability of finding a boson 
in momentum states $\exp(\pm i\pi (N+1)/L)$. 
Note that due to our choice $N=L$ the peaks for different values of 
$N$ approximately coincide at $\pm \pi (1+1/N)\approx \pm \pi$.

\begin{figure}
\begin{center}
\includegraphics[width=0.4 \textwidth ]{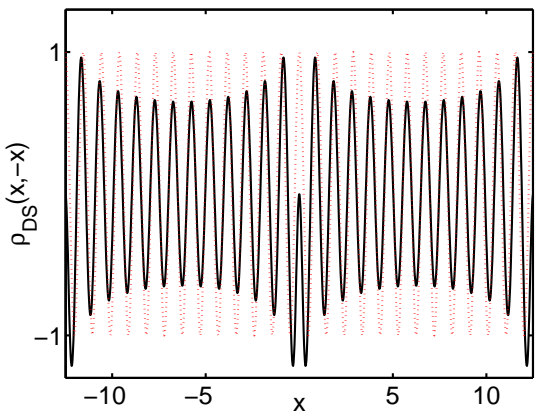}
\caption{ \label{rhoxminx}
(color online) The cross-diagonal $\rho_{DS}(x,-x)$ of the dark "soliton" 
RSPDM (solid black line), displaying long-range oscillatory correlations between 
mirror points $x$ and $-x$, and the cosine function $\cos( 2 k_{peak} x)$
(red dotted line).
}
\end{center}
\end{figure}

The sharp spikes at $k_{peak}=\pm \pi (N+1) /L$ are intimately related to the 
strong correlation between the mirror points $x$ and $-x$. This is 
illustrated in Fig. \ref{rhoxminx} which shows the cross-diagonal 
section of the RSPDM $\rho_{DS}(x,-x)$
and the function $\cos(2k_{peak}x)$ for $N=25$. 
There is evident correlation between $\rho_{DS}(x,-x)$ and 
$\cos(2k_{peak}x)$. 
Because of the symmetry $\rho_{DS}(x,y)=\rho_{DS}(y,x)$, the 
Fourier transform (FT) with respect to $\exp[ik(x-y)]$ reduces to 
FT with respect to $\cos k(x-y)$, which is  $\cos 2 k x$ at 
$y=-x$; hence, from Fig. \ref{rhoxminx} it immediately follows that the 
cross-diagonal behavior of $\rho_{DS}(x,-x)$ induces the 
peaks in the momentum distribution of DS states.
We would like to point out that the correlations $\rho_{DS}(x,-x)$
do not decay, but oscillate, as the separation between points 
$x$ and $-x$ is increased. 

\begin{figure}
\begin{center}
\includegraphics[width=0.4 \textwidth ]{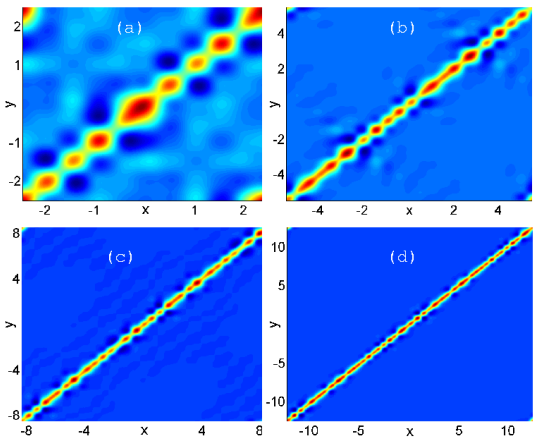}
\caption{ \label{rhorandom}
(color online) The RSPDM of typical eigenstates $\psi_{DE}$
[Eq. (\ref{psiDE})] for $N=7$ (a), 11 (b), 17 (c), and 25 (d). 
}
\end{center}
\end{figure}

In order to gain more insight into the origin of the two sharp 
spikes in the momentum distribution and the related coherence between 
mirror points $x$ and $-x$, it is illustrative to calculate the 
RSPDM and the momentum distribution for eigenstates that are degenerate 
(i.e., that have the same energy) to DS states, but which are 
less restrictive with respect to symmetry of the coefficients 
$a_{m}^{-}$ and $a_{m}^{+}$. 
If the coefficients are chosen as 
$a_{m}^{-}=i \exp(i\theta_m) /\sqrt{2}$ and 
$a_{m}^{+}=-i \exp(-i\theta_m) /\sqrt{2}$,
one obtains a whole class of eigenstates degenerate 
to dark solitons, which have the form 

\begin{eqnarray}
\psi_{DE} & = &
A(x_1,\ldots,x_N)
\left ( \frac{2}{L} \right)^{\frac{N}{2}} \times
\nonumber \\
&&\det_{j,m=1}^{N}[\sin (k_m x_j+\theta_m)],
\label{psiDE}
\end{eqnarray}
where $\theta_m$, $m=1,\ldots,N$ are $N$ phases 
(for $\theta_m=0$, $\psi_{DS}=\psi_{DE}$).

\begin{figure}
\begin{center}
\includegraphics[width=0.4 \textwidth ]{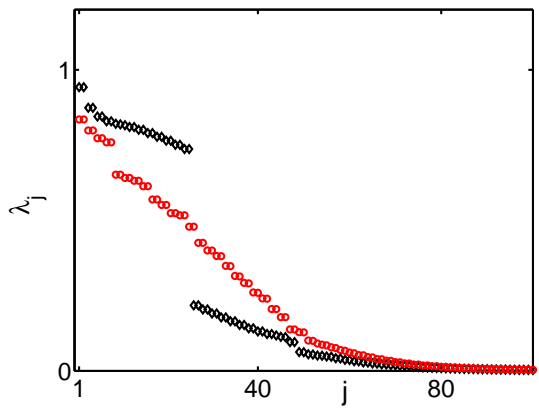}
\caption{ \label{NOocc}
(color online) The occupancies of the NOs for the state 
$\psi_{DS}$ (black diamonds) and a typical state $\psi_{DE}$ (red squares), 
for $N=25$ particles. The sharp drop in the occupancies of $\psi_{DS}$ occurs 
between $\lambda_N$ and $\lambda_{N+1}$. 
}
\end{center}
\end{figure}

Figure \ref{rhorandom} displays contour plots of RSPDMs $\rho_{DE}(x,y)$, 
which corresponds to some typical states $\psi_{DE}$ obtained from 
Eq. (\ref{psiDE}) by randomly choosing $N$ phases $\theta_m$
(with respect to the uniform probability density in $[-\pi,\pi]$).
We see that the SP density for this state, $\rho_{DE}(x,x)$, is not zero at $x=0$, 
which evidently follows from the fact that 
$\sin (k_m x+\theta_m)$ is not an odd function for $\theta_m\neq 0$, 
while $\rho_{DE}(x,x)=\sum_{m=1}^{N}|\sin (k_m x+\theta_m)|^2$. 
Furthermore, we observe that the structure of the RSPDM 
along the $x=-y$ line is absent, that is, there is no coherence
between the mirror points $x$ and $-x$. 
A closely related observation is that the momentum distributions 
of such states do not have a pair of sharp spikes 
which are present in $n_{DS}(k)$; this is illustrated in Fig. \ref{MDdark}(b)
which shows typical momentum distributions $n_{DE}(k)$ for 
$N=11,17$ and $25$. 

\begin{figure}
\begin{center}
\includegraphics[width=0.4 \textwidth ]{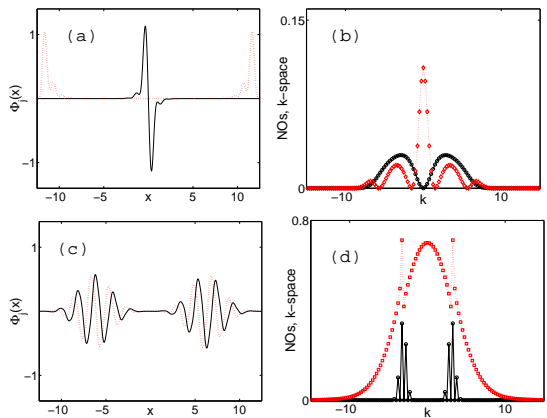}
\caption{ \label{NOs-xandk}
(color online) The NOs of a DS state
in $x$-space (left column) and their power spectra in $k$-space 
(right column) for $N=25$. 
(a) The first $\Phi_1(x)$ (black solid line), and the third $\Phi_3(x)$ (red dotted line) NO. 
(b) The Fourier power spectrum of the first $|\tilde\Phi_1(k)|^2$
(black circles, solid line), and the third $|\tilde\Phi_3(k)|^2$ (red diamonds, dotted line) NO. 
(c) The 24th (red dotted line) and 25th (black solid line) NO in $x$-space.
(d) The momentum distribution (red squares, dotted line) 
in comparison to the contribution from the 
24th and 25th NO: $\sum_{i=24}^{25} 
\lambda_i \tilde|\tilde\Phi_i (k)|^2$ (black circles, solid line).
}
\end{center}
\end{figure}

Besides the RSPDM and the momentum distribution, excited many-body 
eigenstates of interest can be characterized by the corresponding natural 
orbitals (NOs) and their occupancies. 
Figure \ref{NOocc} shows the occupancies of the NOs 
of the state $\psi_{DS}$, and a typical state $\psi_{DE}$, for $N=25$. 
We observe that the occupancies 
are fairly low (less than one) for all NO's, but there is a sharp 
drop in the occupancies after the $25$th NO. We have observed 
such a behavior for other values of $N$ as well. 
In contrast, the occupancies of the NOs corresponding to a typical state $\psi_{DE}$
do not exhibit a sharp drop after the $N$th orbital, but decrease rather smoothly.

Figure \ref{NOs-xandk} illustrates the spatial structure and the Fourier power spectra 
of the NOs corresponding to the DS state for $N=25$. 
The spatial structure of calculated NOs is either symmetric or antisymmetric. 
This is connected to the symmetry $\rho_{DS}(x,y)=\rho_{DS}(-y,-x)$;
due to this symmetry it follows that if some NO is non-degenerate, 
it is either symmetric or antisymmetric; if two NOs are degenerate
(their occupancies are identical), they 
can be superimposed to yield one symmetric and one antisymmetric NO. 
Our numerical study shows that 
the low order (leading) NOs are localized in space, but 
broad in $k$-space; Fig. \ref{NOs-xandk}(a) depicts 
the $x$-space structure, and Fig. \ref{NOs-xandk}(b)  
show the $k$-space structure of the first and the third NO. 
We see that these low order NOs do not 
contribute to the sharp peaks observed in the momentum distribution 
of DS states. Further inspection of the NOs reveals that 
those NOs just on the upper side of the sharp drop in $\lambda_j$
(Fig. \ref{NOocc}) are in fact responsible for the sharp peaks:
Figs. \ref{NOs-xandk}(c) and (d) display the $x$-space and $k$-space 
structure, respectively, of the $24$th and the $25$th NO ($N=25$).
The total momentum distribution (red squares, dotted line in Fig. \ref{NOs-xandk}(d)) 
can be written as $\sum_{i=1}^{\infty} \lambda_i \tilde\Phi_i^* (k) \tilde\Phi_i (k)$;
a contribution to this sum stemming from the $24$th and the $25$th 
NO is shown in Fig. \ref{NOs-xandk}(d) with black solid line. 
Evidently, for this DS state where $N=25$, the $24$th and 
the $25$th NO give rise to the peaks in the momentum distribution.

It is interesting to note that when all phases are chosen to be 
identical but not zero, e.g., if $\theta_m=\pi/2$, then all of the 
fermionic NOs are $\propto \cos (k_m x)$, we again observe a higher degree of 
correlation between mirror points in the RSPDM and peaks in the 
momentum distribution (not shown).

All of the observations above indicate a somewhat smaller degree of order
in the degenerate eigenstates $\psi_{DE}$ than in dark solitons $\psi_{DS}$, 
which follows from the random (disordered) choice of the phases $\theta_m$. 
This is further underpinned in Table \ref{TabEnt}, 
which shows the entropy $S=-\sum_i p_i \log p_i$, where $p_i=\lambda_i/N$, for 
the dark "soliton" states $\psi_{DS}$, and typical $\psi_{DE}$ states. 
The entropy of states $\psi_{DE}$ is systematically larger than 
in the states $\psi_{DS}$. 

\begin{table}
\begin{tabular}{|c|c|c|} 
\hline 
$N$ & $S[\psi_{DS}]$ & $S[\psi_{DE}]$ \\
\hline
11 &  2.90 & 3.21  \\
17 &  3.39 & 3.66  \\
25 &  3.83 & 4.05  \\
\hline
\end{tabular}
\caption{The entropy $S$ of dark "soliton" states, and typical 
$\psi_{DE}$ states for different values of the number of particles $N$.}
\label{TabEnt}
\end{table}

From our observations it follows that the many-body state $\psi_{DS}$ 
contains a distinct component, which can be interpreted as a standing 
wave populating momentum modes at $\pm k_{peak}$. 
In the effective single-particle picture, we see that this component 
give rise to the population of the natural orbitals close to (and including) 
the $N$th NO. However, it should be pointed out that this component 
is fairly small, i.e., it yields small occupation of these effective SP states.

In a similar fashion to the excited $\psi_{DS}$ state, the ground state 
of the TG gas on the ring yields distinct population of the zero-momentum mode \cite{Lenard1964};
in this case, however, the zero-momentum mode is the leading natural orbital, 
and its population is fairly large (it scales as $\sqrt{N}$ 
\cite{Forrester2003}). 
Even though the TG states are not Bose condensed, they can sharply
populate a single momentum mode because bosons do not obey the 
Pauli principle and consequently more than one boson can occupy a single 
momentum state (which is not the case for noninteracting fermions).

It is interesting to note that on the Fermi side of the mapping, 
the momentum distribution of noninteracting fermions $n_F(k)$ is uniform 
up to the Fermi edge (excluding the zero momentum mode at $k_0=0$), 
and does not depend on the randomly chosen phases $\theta_m$:

\begin{equation}
n_{F,DS}(k_m) = n_{F,DE}(k_m) = 
\left\{ 
\begin{array}{ll} 
\frac{1}{2} & \textrm{if $1\leq |m| \leq N$}\\ 
0 & \textrm{otherwise} 
\end{array} 
\right. 
\label{nF}
\end{equation}
Namely, the SP eigenstates $\sin(k_m x+\theta_m)$ are NOs of the 
fermionic system. 
The Fourier power spectrum of each SP state $\sin(k_m x+\theta_m)$
[which determine the fermionic momentum distribution via Eq. (\ref{FMDNOs})],
does not depend on on the phase $\theta_m$. 
Each fermionic NO $\sin(k_m x+\theta_m)$ can be written 
as a superposition of two plane waves 
$\sin(k_m x+\theta_m)=(e^{ik_m x+i\theta_m}-e^{-ik_m x-i\theta_m})/2i$. 
Evidently, the mean value of the momenta pointing in the positive (negative) 
direction is $\pi (N+1) /L$ [$-\pi (N+1) /L$, respectively], that is, 
it is identical to $k_{peak}$. 
When the fermionic states are mapped to the TG states, 
a wave function component which distinctively populates momentum 
modes at $\pm k_{peak}$ can appear. This occurs when the 
phases $\theta_m$ act coherently, i.e., it is evident that 
a random choice of the phases $\theta_m$ destroys the 
observation of the two spikes connected with this component.

\begin{figure}
\begin{center}
\includegraphics[width=0.4 \textwidth ]{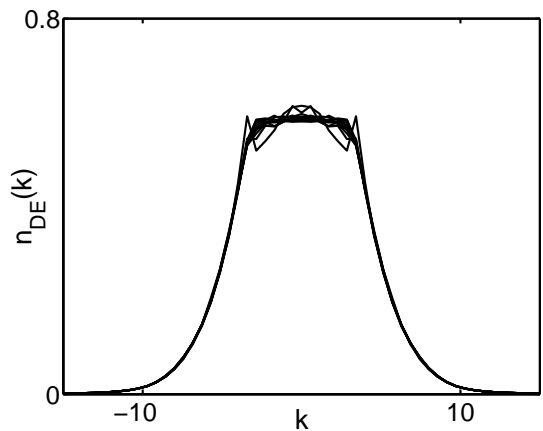}
\caption{ \label{figEns}
Momentum distributions for $10$ different states $\psi_{DE}$
[Eq. (\ref{psiDE})] chosen at random, by randomly choosing 
$10$ sets $\{ \theta_m\ |\ m=1,\ldots,N \}$ of phases  
(see text for details). 
}
\end{center}
\end{figure}

Before closing this section we should say that in all our numerical
calculations, the phases of the states $\psi_{DE}$ were chosen at random
(with respect to the uniform probability density in $[-\pi,\pi]$). 
A random choice of the phases yields a typical state $\psi_{DE}$
in the sense that one-body observables, such as the momentum distribution, 
of typical states approximately coincide. 
In order to verify this assumption Figure \ref{figEns} displays momentum distributions 
for 10 eigenstates $\psi_{DE}$ ($N=11$ particles), obtained by 10 randomly chosen 
configurations $\{ \theta_m\ |\ m=1,\ldots,N \}$ of the phases. 
The momentum distribution only slightly varies from case to case with 
one exception that exhibits $2$ (relatively small) dark solitonic spikes. 
Exceptions from the typical behavior will be harder to see
for larger values of $N$, because in this case the parameter space 
spanned by $N$ phases $\theta_m$ is larger, and it is harder
to correlate the phases by chance, which could yield characteristic solitonic
spikes in the momentum distribution. 
Hence, we can conclude that our observations regarding the 
class of states $\psi_{DE}$ from Eq. (\ref{psiDE}) hold for practically all 
of these states in the sense stated above.

\section{DS states in a parity invariant well-shaped potential}

\begin{figure}
\begin{center}
\includegraphics[width=0.4 \textwidth ]{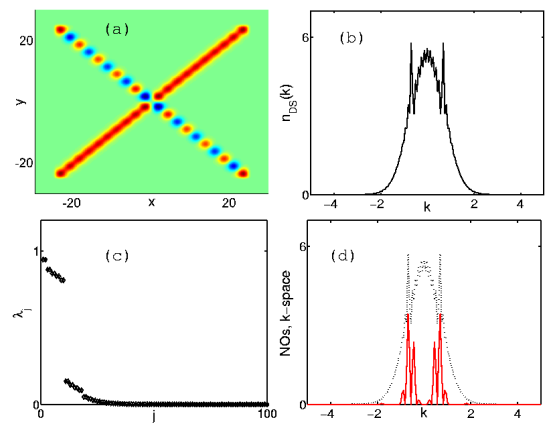}
\caption{ \label{box}
(color online) The RSPDM (a), momentum distribution (b), and 
occupancies of the NOs (c) for a DS state ($N=10$) 
in a parity-invariant well-shaped potential.
(d) The momentum distribution (black dotted line) 
in comparison to the contribution from the 
9th and 10th NO: $\sum_{i=9}^{10} 
\lambda_i |\tilde\Phi_i (k)|^2$ (red solid line).  
}
\end{center}
\end{figure}

The concept of dark "solitons" can be extended to various
types of parity-invariant potentials (e.g., see \cite{Buljan2006}). 
DS states are found in harmonic confinement \cite{Busch2003}
periodic lattices \cite{Buljan2006}, well-shaped potentials \cite{Buljan2006},
and so-forth. 
In Ref. \cite{Buljan2006} it was shown that by
parity invariant filtering of the many-body wave function, 
one could in principle excite the TG gas into a 
DS state. 
Let us compare the RSPDM and the momentum distribution 
of DS states on the ring, and in a parity invariant potential 
$V_c(x)=V_c^0 \{ 2 + \sum_{i=1,2} (-)^{i+1}\tanh x_w(x+(-)^i x_c) \}$ 
($V_c^0=15$, $x_w=8$, and $x_c=25$).
In such potential, DS states are constructed by 
populating the first $N$ {\em odd} SP eigenstates 
on the Fermi side of the map.
Figure \ref{box}(a) displays the RSPDM, while Fig. \ref{box}(b) displays the 
momentum distribution of such an excited eigenstate for $N=10$. 
We clearly observe that the structure of the RSPDM and the 
momentum distribution is similar to that of DS states
on the ring; the RSPDM has off-diagonal mirror-point correlations,
while the momentum distribution has two sharp spikes. 
Furthermore, Fig. \ref{box}(c) displays the occupancies of the NOs,
which clearly exhibit a large and sudden drop after the $N$th NO. 
Fig. \ref{box}(d) shows the contribution from the 
$(N-1)$th and the $N$th NO to the momentum distribution: 
$\sum_{i=9}^{10} \lambda_i \tilde\Phi_i^* (k) \tilde\Phi_i (k)$;
clearly these NOs are responsible for the peaks in the momentum 
distribution. 

The observations presented in Fig. \ref{box} suggest that 
the behavior of the one-body observables of DS states, such as 
the two sharp spikes in the momentum distribution and correlation 
between the mirror points in the RSPDM can be found 
in various types of parity-invariant potentials.

\section{Connection to incoherent light}

We would like to point out that the behavior of incoherent light 
in linear \cite{Turunen} and nonlinear \cite{Moti,SpatSolitons,Cohen2005,Picozzi} 
optical systems has many similarities to the behavior of interacting 
(partially condensed or non-condensed) Bose gases 
\cite{Buljan2006,Picozzi,Naraschewski1999,Buljan2005}. 
The dynamics of incoherent light in nonlinear systems attracted considerable 
interest in the past decade since the first experiments on incoherent solitons 
\cite{Moti} in noninstantaneous nonlinear media were conducted. 
A number of important results were obtained (for a review e.g., 
see Ref. \cite{SpatSolitons}) since then. Among the recent results 
one finds, e.g., the experimental observation of incoherent solitons 
nonlinear photonic lattices \cite{Cohen2005}, and thermalization 
of incoherent nonlinear waves \cite{Picozzi}. 
We believe that many of the phenomena observed with incoherent light in optics 
\cite{Moti,SpatSolitons,Cohen2005,Picozzi} can find its counterpart
in the context of Bose gases. 

In Ref. \cite{Buljan2006} it has been pointed out that there is mathematical 
relation between the propagation of partially spatially incoherent light 
in {\em linear} 1D photonic structures and quantum dynamics of a TG gas. 
More specifically, the correlation functions describing incoherent 
nondiffracting beams in optics \cite{Turunen}
can be mapped \cite{Buljan2006} to DS states. 
However, it should be emphasized that the spatial power spectrum of 
these incoherent beams corresponds to the momentum distribution of 
noninteracting fermions, i.e., it profoundly differs from the momentum 
distribution of DS states in a TG gas discussed here.

\section{Summary}

We have employed a recently obtained formula \cite{Pezer2007} to 
numerically calculate the RSPDM correlations, natural orbitals and 
their occupancies, and the momentum distribution of dark
"solitons" in a TG gas. We have found that these excited eigenstates 
of a TG gas have characteristic shape of the momentum 
distribution, which has two distinguished sharp spikes; while most of 
the paper is devoted to the ring geometry, where the spikes are 
located at $k_{peak}=\pm\pi (N+1)/L$ ($N$ is the number of particles and 
$L$ is the length of the ring), we have shown results which 
suggest that such behavior is general for DS states in parity invariant potentials. 
It has been shown that the spikes in the momentum distribution are 
closely connected to the cross-diagonal oscillatory long-range 
correlations between mirror points ($x$ and $-x$) in the RSPDM. 
This behavior of DS states follows from the fact that they are specially 
tailored; in the ring geometry, it has been shown 
that the two spikes and a special form of spatial coherence are lost for 
most eigenstates that are degenerate to DS states.

\section{Acknowledgments}

This work is supported by the Croatian Ministry of Science
(grant no. 119-0000000-1015).

\end{document}